\documentclass{article}
\usepackage{iclr2027_conference,times}
\iclrfinalcopy
\usepackage{amsmath,amssymb,booktabs,graphicx,microtype,hyperref,url,array}
\hypersetup{hidelinks}
\graphicspath{{figures/}}
\title{Perceptual Reality Transformer:\\What Must an Illustration Preserve?}
\author{Baihan Lin\\\texttt{doerlbh@gmail.com}}

\author{Baihan Lin \\
\texttt{baihan.lin@mssm.edu} \\
}

\begin{document}
\maketitle
\lhead{Preprint}
\begin{abstract}
How can models help people communicate unusual perceptual experiences without changing what they mean? A recognizable image is only part of the answer: accounts also express vividness, duration, uncertainty, and emotion. We introduce Perceptual Reality Transformer (PRT), an evidence-linked workflow and descriptive atlas that retain the source account alongside representations and generated illustrations. Three studies examine successive parts of this transformation. In 3,145 public Ganzflicker records, frozen neural representations predict complex content well (AUROC 0.887--0.894), but vividness, persistence, and unpleasantness are less recoverable; separate imagery/context responses improve vividness prediction relative to narrative embeddings alone. In 672 synthetic contrasts, similarity judgments depend strongly on wording, while exploratory probes recover some distinctions that cosine handles poorly. Finally, 72 matched illustrations from 24 reports compare direct prompting, evidence in prose, and structured PRT. The automated coverage metric does not demonstrate a structured-format advantage; near-ceiling scores, an apparent judge error, and an analogy-to-object scoring ambiguity limit its interpretation. Together, the studies motivate separating missing information, inaccessible meaning, and an inadequate evaluation criterion. The resulting atlas makes these choices inspectable, without treating generated images as measurements of private perception.
\end{abstract}

\section{Introduction}
For someone living with a mental illness or neurological disorder, describing an unusual perceptual experience can be difficult even when it is vivid. An illustration might help a family member or clinician understand a visual disturbance. The aim is communication: give another person something concrete to discuss while preserving the account's qualifications and context. A diagnosis alone cannot specify what an individual sees or how the experience feels.

\begin{figure}[t]
\centering\includegraphics[width=\linewidth]{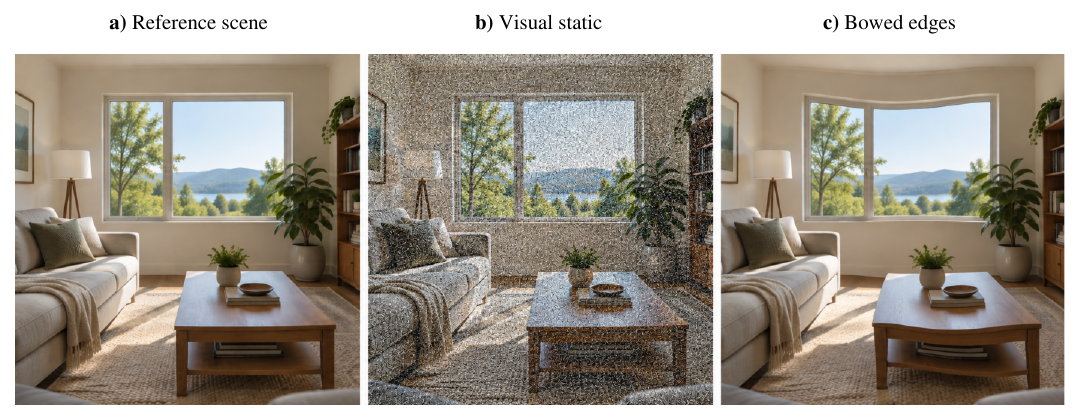}
\caption{\textbf{The communication goal.} (a) A reference scene; (b) visual static; (c) bowed window and table edges. These existing, author-specified generated illustrations make visible changes easy to discuss while retaining an ordinary scene. They are not participant reconstructions or experimental treatment conditions; incidental scene changes prevent a controlled pixel comparison. Time course, certainty, and distress remain qualities to communicate alongside the image.}
\label{fig:visual}
\end{figure}

Figure~\ref{fig:visual} illustrates the attraction and limit of this idea. Visual static or bent edges are easier to discuss against an everyday scene. But a still image does not say whether the effect lasted a moment, felt distressing, or was only a tentative impression. Even a compelling depiction can omit what matters most to the person describing it.

Language and vision-language models can encode an account, retrieve similar accounts, or generate an image. Each operation makes a different promise. A narrative may omit an attribute. An embedding may retain it while a similarity score overlooks it. An image can satisfy a checklist whose wording already changed the source's meaning. Treating these as one generic ``fidelity'' problem makes it difficult to know what to improve.

We introduce \emph{Perceptual Reality Transformer} (PRT), a source-linked evidence representation, prompt compiler, and descriptive atlas using an existing image generator. It is not a newly trained Transformer. We investigate three questions: \emph{what reported qualities are recoverable from an account; what meaning survives a representation's readout; and what an image comparison actually establishes.} The contribution is an implemented communication workflow together with three audits that expose different limits of recognizable content as a proxy for experience.

We use public Ganzflicker-induced visual experiences \citep{reeder2022}, which provide real accounts and separate questionnaire responses. Clinical communication motivates the work, but these induced-experience records do not establish clinical validity. Our synthetic contrasts and AI-authored image annotations remain separate from the original observations. The studies preserve negative and ambiguous findings because those findings determine what a reader can conclude from an attractive image.

\section{PRT: preserve the account through the transformation}\label{sec:prt}
\paragraph{A source-linked record.}
PRT takes an account and evidence annotations, then returns the original account, annotations, compiled prompt, and generated image with provenance. Each annotation includes an exact source span, an interpretation for illustration, and a status such as reported, uncertain, memory, analogy, or interpretation. ``Like when you pour milk into coffee,'' for example, can describe billowing appearance without reporting a cup of coffee. Source-span validation rejects annotations without matching text; it checks traceability, not whether an interpretation is correct. Annotations here are AI-authored and fixed before generation, without independent human validation.

The compiler asks the generator to preserve supported visible attributes and retain qualifications. It does not fill gaps from a diagnosis. Temporal and nonvisual qualities remain in the account. These are model instructions, not output guarantees. The study requests an image in every condition and does not evaluate abstention. The atlas stores 3,145 accounts with search, questionnaire filters, and comparisons; 24 accounts have matched illustrations. ``Atlas'' denotes this inspectable collection, not a taxonomy of disorders (Appendices~\ref{app:atlas} and~\ref{app:atlas-views}).

\paragraph{A concrete transformation.}
Report 5403 says, ``The flickers would change speed and the dark areas would go counter-clockwise.'' Its frozen evidence record links ``dark areas'' to ``dark regions within a flickering field,'' with status \emph{reported}. The compiler includes this record and the full account; the visible-detail question asks whether dark regions appear. Changing speed and rotation remain in the account but are not scored. This example makes the boundary explicit: PRT organizes selected claims for illustration; it does not reconstruct every aspect of a report.

\paragraph{Separating evidence from its presentation.}
Let $r$ be the account, $e$ its fixed evidence record, $\pi$ the shared preservation policy, and $G$ the image backend. We compare
\begin{align}
I_D &\sim G(\text{task},r), \nonumber\\
I_E &\sim G(\text{task},\pi,\operatorname{prose}(e),r),\\
I_S &\sim G(\text{task},\pi,\operatorname{fields}(e),r).\nonumber
\end{align}
Every condition includes the verbatim account. Evidence prose ($E$) and structured PRT ($S$) include the same statuses, spans, and interpretations; only presentation differs. Comparing $S$ with direct prompting ($D$) instead measures the combined effect of annotation, instructions, and structure. The record makes these choices inspectable even when an output looks convincing. Whether this helps contributors or readers remains an empirical question.

\section{Related work}
Ganzflicker research connects imagery capacity with induced experience \citep{reeder2022,chkhaidze2026}; MOSAIC applies topic modeling and language models to stroboscopic reports \citep{beaute2026}. Our question is which reported dimensions remain accessible as accounts become representations and illustrations, not whether these accounts define a new perceptual taxonomy. MUSE maps emphasize the embodied context of unusual sensory experience \citep{melvin2021}, while work on VLM simulations of low vision distinguishes plausible transformations from intended perception \citep{natalie2025}. These motivate communication without supplying participant-endorsed image targets for our study.

Sentence encoders \citep{reimers2019,song2020} and CLIP \citep{radford2021} provide reusable representations. Compositionality diagnostics show why object agreement can miss relations and qualifications \citep{yuksekgonul2023,quantmeyer2024}, but diagnostic construction can itself introduce shortcuts \citep{hsieh2023}. Encoded information also differs from what a particular similarity or alignment rule exposes \citep{koishigarina2026,lu2026}. We retain that distinction: a similarity failure does not prove an attribute was erased, and a successful probe does not show that a generator preserves it.

\section{Study 1: which reported qualities are recoverable?}\label{sec:prediction}
A narrative naming a geometric pattern may say little about how vivid, persistent, or unpleasant it was. We test recovery of these qualities using separate questionnaire responses, rather than model-invented labels.

\paragraph{Data and evaluation.}
We retain consented Ganzflicker records with reported ages 18--100 and descriptions of 5--400 words: 3,145 records, comprising 3,134 normalized-text groups. Targets distinguish complex objects/environments, at least moderate vividness, persistence of several seconds or longer, and unpleasant affect. Ambiguous or missing targets are excluded per task. These are heterogeneous self-reports without a validated person identifier; records are not necessarily distinct people. Appendix~\ref{app:data} gives exact eligibility and label mappings.

Text-hash groups define approximately 60/20/20 training/development/test splits (1,897/629/619 records); inspected pilot groups stay in training and exact duplicates cannot cross splits. Valid test sizes are 537/474/450/619 for the respective targets. We compare TF--IDF, 24-component NMF \citep{lee1999}, length/style features, and frozen MiniLM, MPNet, and CLIP text, plus a separate imagery/context input channel. Neural models share 32-word chunk pooling. Logistic heads are selected by development AUROC, with preprocessing fitted on training data. AUROC measures ranking, with chance at 0.5; 95\% intervals use 500 group-bootstrap resamples, conditional on fitted models. Unknown contributor overlap remains a limitation. Full protocols and additional metrics are in Appendices~\ref{app:prediction-protocol} and~\ref{app:metrics}.

\begin{figure}[t]
\centering\includegraphics[width=\linewidth]{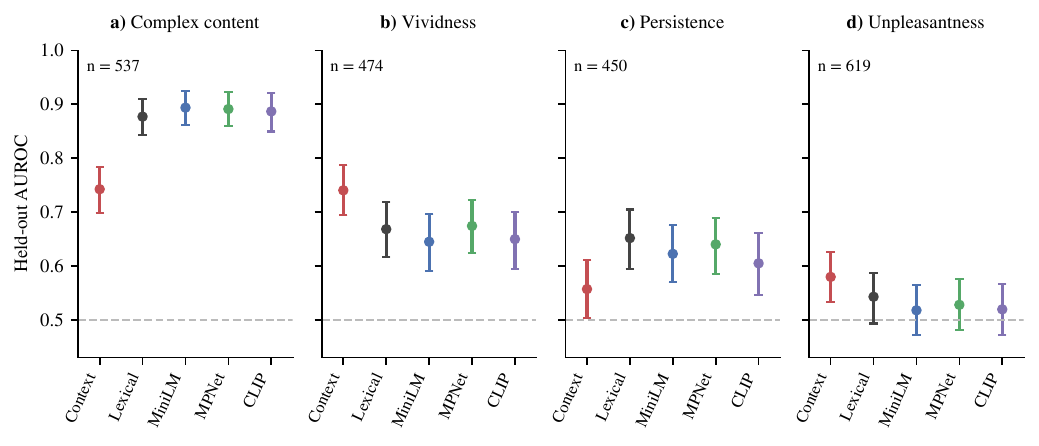}
\caption{\textbf{Content is easier to predict than other reported qualities.} Held-out AUROC for (a) complex content, (b) vividness, (c) persistence, and (d) unpleasantness, with 95\% group-bootstrap intervals. Context uses separate imagery and recording-condition responses; other models use the narrative. These are original questionnaire targets, not model-generated labels.}
\label{fig:channels}
\end{figure}

\paragraph{Content prediction is not a complete account.}
Figure~\ref{fig:channels} shows strong complex-content AUROC for MiniLM (0.894), MPNet (0.891), and CLIP text (0.887), versus TF--IDF (0.877). The paired MiniLM--TF--IDF difference is 0.017, with 95\% interval $[-0.013,0.048]$, so these results do not establish a reliable ranking advantage. At threshold 0.5, neural content balanced accuracy is 0.758--0.771 versus 0.590 for TF--IDF, with lower neural Brier scores: similar AUROC does not imply similar probability predictions (Table~\ref{tab:metrics}).

Narrative neural AUROC is lower for vividness (0.645--0.674), persistence (0.605--0.640), and unpleasantness (0.517--0.528); TF--IDF is competitive. This is compatible with the source prompt emphasizing visible content. Questionnaire responses may prime descriptions or share vocabulary with them. Strong content prediction is therefore a measurement check, not evidence that the full experience has been captured.

\paragraph{Reporting channel matters.}
Imagery/context responses predict vividness at 0.740 AUROC. Exploratory MPNet-plus-context reaches 0.733: better than MPNet text alone, but a difference of $-0.007$ $[-0.050,0.039]$ from context alone. Appendix Figure~\ref{fig:sensitivity} separates the text-only, context-only, and combined-input comparisons. Context supplies additional inputs, so it is not a text-only competitor. Their value is consistent with known imagery associations \citep{reeder2022}; it does not establish a novel psychiatric predictor.

A temporal group split retains high neural content AUROC (0.889--0.911) and weak affect AUROC (0.508--0.539). Lexical masking reduces content AUROC to 0.859. NMF retains some content information (0.804) but little affect prediction (0.487). These supporting results, including context comparisons, topic loadings, and permutation checks, appear in Appendix Figures~\ref{fig:sensitivity} and~\ref{fig:topics} and Appendices~\ref{app:metrics} and~\ref{app:topics}. They are within-source checks, not external replication.

A weak score can reflect omitted information, noisy responses, or an unsuitable representation. The practical implication is to retain separate reporting channels, rather than ask one content embedding to stand for the whole experience.

\section{Study 2: does similarity preserve the meaning?}\label{sec:contrasts}
``I saw a pattern'' and ``I imagined a pattern'' share content but make different claims. Unlike the questionnaire study, controlled sentences let us hold the relevant information in the text and specify which distinction should survive.

\paragraph{Controlled contrasts.}
We construct 672 AI-authored trials: 24 visual contents $\times$ seven operators $\times$ two polarities $\times$ two distractor forms. Operators concern certainty, negation, duration, motion, location, control over appearance, and seeing versus imagining. Each query $q$ is compared with a faithful paraphrase $p$ and a same-content, meaning-changing alternative $n$. The minimal distractor changes little wording; the paraphrased distractor expresses the opposite meaning in alternative wording. A different-content control checks content discrimination. These are template diagnostics, not human-validated accounts (Appendix~\ref{app:contrasts}).

For frozen encoder $f$ and cosine similarity $s$, which compares vector directions, we score
\begin{equation}
m(q,p,n)=s(f(q),f(p))-s(f(q),f(n)).\label{eq:margin}
\end{equation}
Positive margins count as correct, with half credit for ties. This tests which meaning the similarity rule favors, not every distinction the representation could support. TF--IDF fitted to diagnostic strings provides a lexical reference.

\begin{figure}[t]
\centering\includegraphics[width=\linewidth]{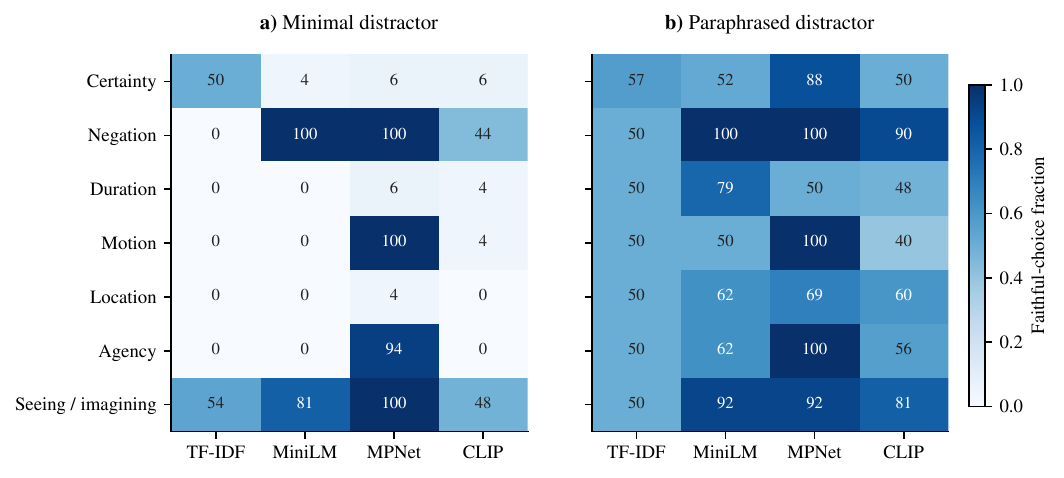}
\caption{\textbf{Diagnostic wording changes the conclusion.} Each cell is faithful-choice accuracy (percentage) over 24 contents and two polarities. (a) Minimally changed opposite-meaning distractor. (b) Paraphrased opposite-meaning distractor. These are synthetic text-to-text tests. Results are conditional on the seven template families; the 672 trials are not 672 independent human judgments.}
\label{fig:contrasts}
\end{figure}

\paragraph{Wording changes the apparent failure.}
Figure~\ref{fig:contrasts} shows large differences between distractor forms. CLIP text struggles with several minimal changes, yet reaches 89.6\% on paraphrased negation. MiniLM's negation score is perfect in these templates, while its minimal duration, motion, location, and agency scores are zero. MPNet handles motion and negation well in both forms but remains sensitive to duration and location wording. These outcomes resist blanket claims that an encoder ``understands negation'' or ``does not preserve experience.''

The lexical reference fails on most minimal contrasts: the wrong option deliberately repeats the query's wording. The alternative distractor form is therefore essential. Neural models generally do better on the different-content control, but that does not validate the operator test. Benchmark shortcuts \citep{hsieh2023} can make an attractive diagnostic exaggerate or misattribute a failure.

\begin{figure}[t]
\centering\includegraphics[width=\linewidth]{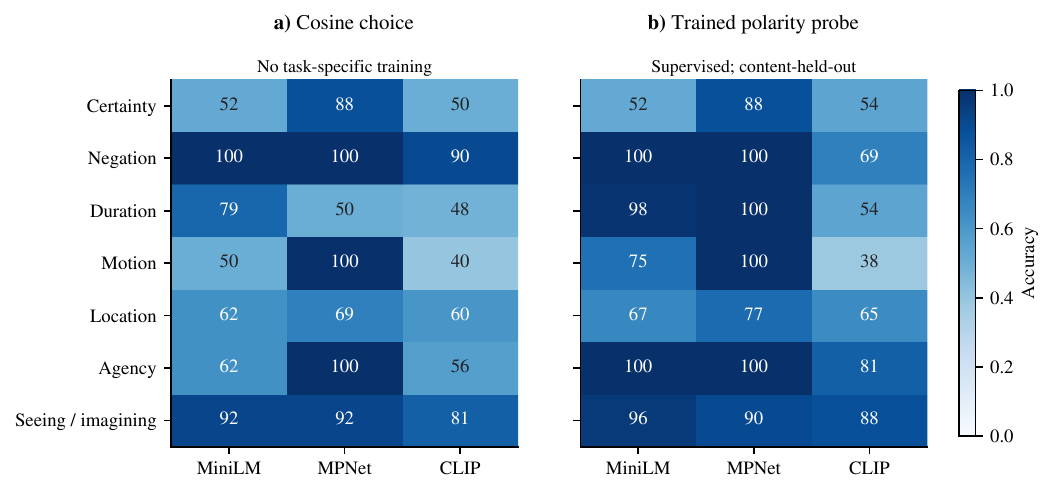}
\caption{\textbf{An exploratory readout audit.} (a) Cosine faithful-choice accuracy with paraphrased distractors. (b) Content-held-out polarity classification accuracy from trained probes. Cells show percentages for MiniLM, MPNet, and CLIP text. These are different tasks and supervision regimes, not a matched improvement comparison. Their contrast shows why poor cosine ordering alone cannot establish that no usable attribute information remains.}
\label{fig:readout}
\end{figure}

\paragraph{Recoverability differs from similarity.}
Exploratory operator-specific linear probes use four content-held-out folds, training on anchor wording and testing on paraphrases. A probe is a simple classifier on unchanged encoder outputs. Average polarity accuracy across operators is 83.9\% for MiniLM, 93.5\% for MPNet, and 64.0\% for CLIP text. Recovery is uneven: CLIP's motion probe reaches 37.5\%, its agency probe 81.3\% (Figure~\ref{fig:readout}; Appendix~\ref{app:readout}).

Probes and cosine tests use different tasks and supervision; these are not matched improvements. Nevertheless, poor similarity ordering alone cannot establish absence of usable attribute information. An atlas should evaluate the operation its readers use: retrieval can return the wrong qualification even when a supervised classifier can recover it. These small, operator-specific tests do not establish a neural mechanism or generator fidelity.

\paragraph{Why an accessible attribute can lose under cosine.}
A simple constructed example makes the distinction precise. Let $u,u'$ be unit content vectors with $u^\top u'=\rho<1$, and let the last coordinate encode a binary qualification at scale $\epsilon>0$. Set $q=(u,\epsilon)$, its faithful paraphrase $p=(u',\epsilon)$, and the opposite qualification $n=(u,-\epsilon)$. All have norm $\sqrt{1+\epsilon^2}$, so
\begin{equation}
s(q,p)-s(q,n)=\frac{\rho-1+2\epsilon^2}{1+\epsilon^2}.\label{eq:readout-example}
\end{equation}
Cosine prefers the wrong qualification whenever $2\epsilon^2<1-\rho$, although a linear readout of the last coordinate recovers its sign exactly. In plain language, a change in content wording can outweigh a fully retained but weakly weighted qualification. This is an illustrative geometric counterexample, not a fitted explanation of our encoders. It shows why a failed similarity test and a successful probe can coexist, and why evaluating the deployed readout matters.

\section{Study 3: what do an image and its score preserve?}\label{sec:controlled}
The first two studies motivate preserving qualifications but do not evaluate image generation. Here we compare visible-attribute coverage across the three prompt conditions, then inspect whether the metric asks and answers the right question.

\paragraph{Matched generation.}
We select 24 unique test groups with 12--70-word accounts by fixed salted text hash, before annotation or generation. There are 74 source-linked annotations and 69 fixed questions: 66 supported visible details and three narrow unsupported-literalization probes. The initially frozen plan requested two outputs per condition (144 calls). Generation paused after 62 outputs for budget reasons; an amendment added 22 missing-condition calls. The primary analysis takes the earliest output per report/condition: 72 matched images, with 12 extra outputs retained for sensitivity. Selection never uses image quality. Prompts and questions are unchanged; this is a budget-amended pilot, not the completed original repeat design.

Embedded output metadata identifies the image backend as \texttt{gpt-image}, version 2.0. Its interface does not expose exact weights, seeds, or internal prompt processing. We match calls, not latent draws or compute. The separate Qwen2.5-VL-3B-Instruct judge \citep{qwen2025vl} sees images and fixed questions without condition labels, generation prompts, or expected answers. It passes 32 elementary fixtures, which check pipeline behavior rather than complex-scene accuracy. Appendix~\ref{app:image-protocol} gives the full protocol; Appendix~\ref{app:repro} records execution details.

\paragraph{Coverage and uncertainty.}
Let $y_{iak}$ be the judge's answer to supported question $k$ for report $i$ and arm $a$, with $m_i$ supported questions. We compute
\begin{equation}
c_{ia}=\frac{1}{m_i}\sum_{k=1}^{m_i}\mathbf{1}\{y_{iak}=\mathrm{yes}\},\qquad
\widehat\Delta_{S-D}=\frac{1}{24}\sum_{i=1}^{24}(c_{iS}-c_{iD}).\label{eq:coverage}
\end{equation}
Uncertain/unparseable answers are not confirmed. Equal report weighting prevents accounts with more questions from dominating; changing one answer changes an arm's mean by $100/(24m_i)$ percentage points. The primary comparison is structured PRT minus evidence-matched prose; 95\% intervals use 10,000 paired report-bootstrap resamples and tests use 100,000 sign flips. The unit is 24 accounts, not 72 independent people. Questions omit time course, emotion, and contributor endorsement; the three literalization probes are not a comprehensive hallucination measure.

\begin{figure}[t]
\centering\includegraphics[width=\linewidth]{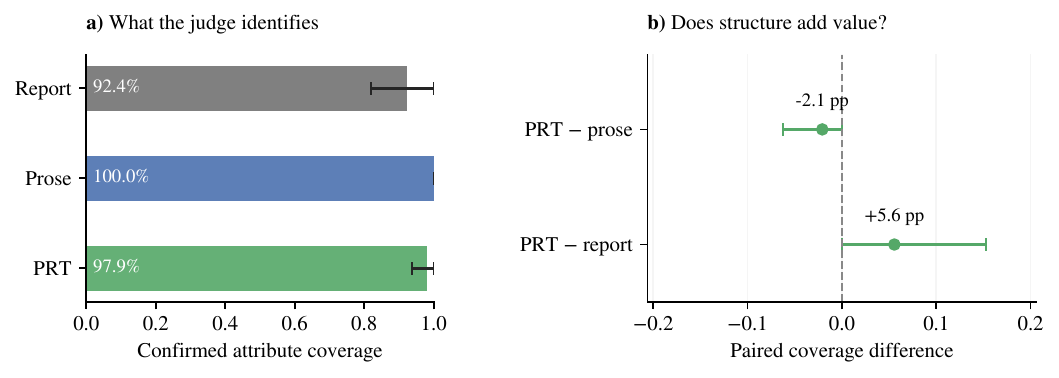}
\caption{\textbf{Visible-attribute coverage under a fixed generation budget.} (a) Mean confirmed coverage from the separate automated judge, weighting reports equally. (b) Paired differences for structured PRT versus evidence-matched prose (primary) and versus the original report (secondary). Intervals are 95\% report-bootstrap intervals. Each report contributes one image per condition. These are model judgments of source-linked visible details, not human ratings of experiential fidelity.}
\label{fig:controlled-results}
\end{figure}

\paragraph{No demonstrated structured-format advantage.}
Mean coverage is 92.4\% direct, 100.0\% prose, and 97.9\% structured PRT (Figure~\ref{fig:controlled-results}; Table~\ref{tab:controlled}). PRT minus prose is $-2.1$ percentage points (95\% interval $[-6.2,0.0]$, $p=1.000$); PRT minus direct is $5.6$ points ($[0.0,15.3]$, $p=0.497$). Across 207 judgments there are no uncertain or unparseable answers. All three literalization probes are negative in every arm, making them uninformative about comparative performance. The all-output sensitivity gives identical macro scores.

Prose receives yes on all 66 supported questions; PRT has one judged miss and direct has three. The prose bootstrap interval collapses because every observed report scores one, not because population performance is known perfectly. Near-ceiling coverage can reflect permissive questions or an insensitive judge. Traceability is implemented; superior generation fidelity is not demonstrated.

\begin{figure}[t]
\centering\includegraphics[width=\linewidth]{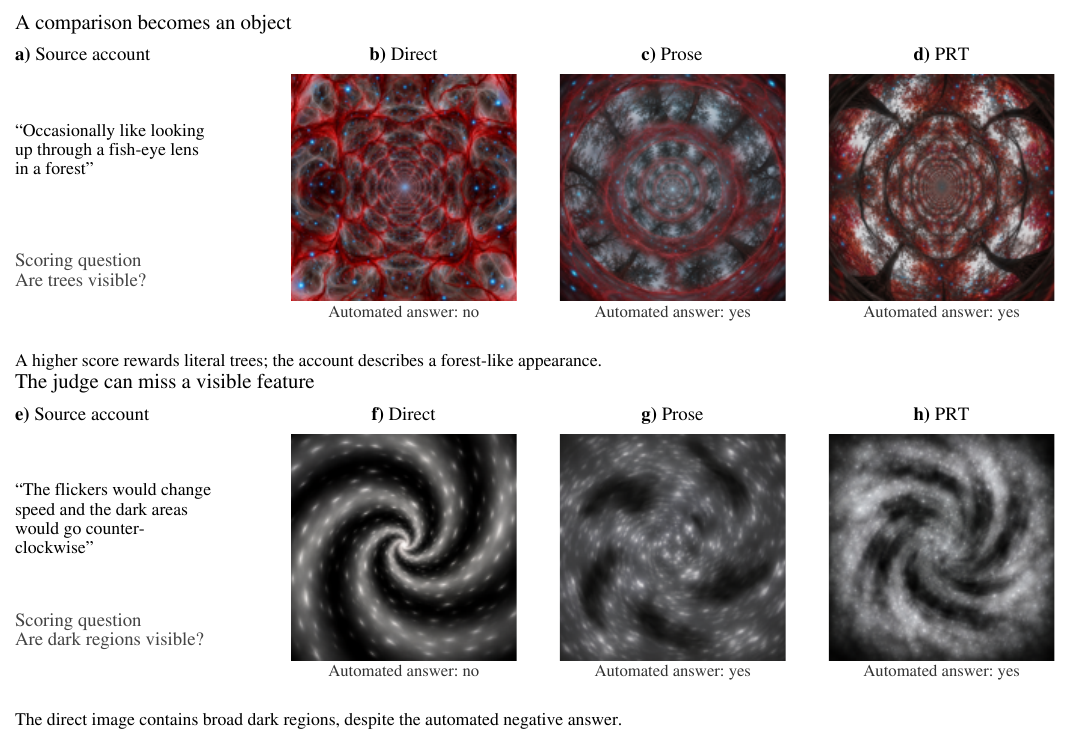}
\caption{\textbf{Inspect the criterion before declaring a better image.} (a--d) Report 3659: (a) a source excerpt uses a forest analogy; (b) direct, (c) prose, and (d) PRT outputs receive different answers to a question requiring visible trees. Literal trees satisfy that question, but are not automatically a more faithful interpretation of the analogy. (e--h) Report 5403: (e) the account mentions dark areas; the judge answers no for (f) direct despite visible dark spiral regions, and yes for (g) prose and (h) PRT. Question wording is shortened in the figure; the exact questions and full accounts are saved. These post hoc diagnostic cases illustrate annotation and judge concerns, not a performance ranking or contributor endorsement.}
\label{fig:controlled-examples}
\end{figure}

\paragraph{A higher score can answer the wrong question.}
Figure~\ref{fig:controlled-examples} shows two post hoc diagnostic cases, retaining the earliest output in every arm. First, report 3659 describes a view ``like looking up through a fish-eye lens in a forest.'' The frozen question requires visible trees. Prose and PRT depict trees and score yes; direct scores no. Literal trees satisfy that criterion, but can over-specify an analogy describing a pattern. Both evidence arms share the same AI-authored interpretation, so their agreement does not validate it. This flags ambiguity, not proof that direct is better; the frozen question and scores remain unchanged.

Second, report 5403 mentions dark areas. All three outputs visibly contain dark regions, but the judge answers no for direct. We retain that apparent false negative and flag it. These problems require different checks: source-to-question interpretation and image-judge accuracy. Improving a score before checking both could reward a changed meaning. Neither case is an unbiased estimate of error prevalence or contributor endorsement.

\paragraph{Where the aggregate difference comes from.}
Table~\ref{tab:report-decomposition} decomposes the unchanged primary scores. PRT and direct tie on 22 reports; the entire $5.56$-point difference comes from the dark-region case ($4.17$ points) and forest case ($1.39$ points). Thus the two diagnostic concerns affect every positive report-level contribution, not merely peripheral examples. Against prose, PRT ties on 23 reports and scores lower on one. This post hoc decomposition explains the observed mean; it neither corrects the judge nor estimates performance after removing difficult cases.

\begin{table}[t]
\centering
\caption{\textbf{The aggregate contrast rests on few reports.} Within-report confirmed coverage for every report with unequal arm scores. ``Contribution'' is that row's contribution to the overall PRT-minus-direct difference in Equation~\ref{eq:coverage}, in percentage points. The remaining 21 reports have perfect automated coverage in all arms. These are original judge outputs, not adjudicated fidelity.}\label{tab:report-decomposition}
\begin{tabular}{lrrrrr}
\toprule
Report & Questions & Direct & Prose & PRT & Contribution\\
\midrule
1591 & 2 & 50.0\% & 100.0\% & 50.0\% & +0.00 pp\\
3659 & 3 & 66.7\% & 100.0\% & 100.0\% & +1.39 pp\\
5403 & 1 & 0.0\% & 100.0\% & 100.0\% & +4.17 pp\\
Other 21 reports & --- & 100.0\% & 100.0\% & 100.0\% & 0.00 pp\\
\bottomrule
\end{tabular}

\end{table}

\section{Discussion and limitations}\label{sec:atlas}
\begin{figure}[t]
\centering\includegraphics[trim=90bp 20bp 90bp 24bp,clip,width=\linewidth]{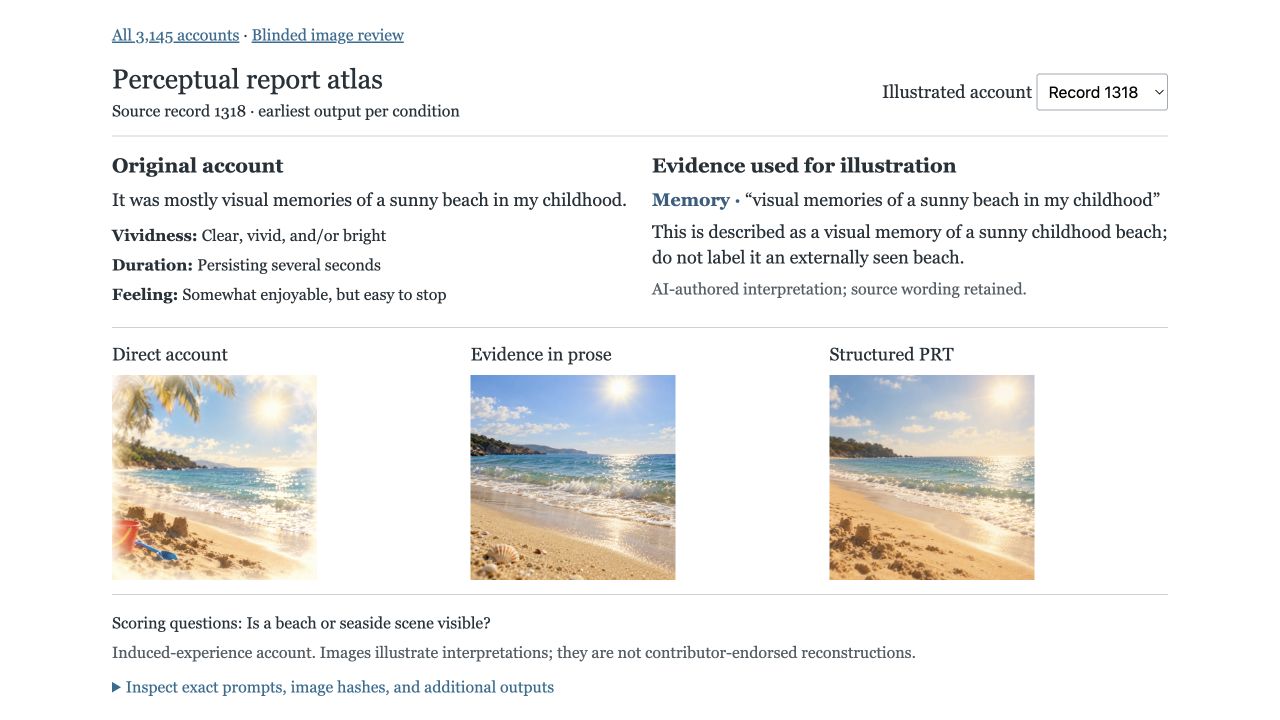}
\caption{\textbf{The working atlas: inspect an illustration against its source.} Screenshot of the offline record view for account 1318, selected to illustrate a memory qualification. The account and questionnaire responses appear beside the evidence record; below are the earliest outputs from all three conditions. The memory status stays explicit even though each image depicts a beach. The expandable control exposes exact prompts, image hashes, and retained repeats. This is an interface demonstration, not a reconstruction endorsed by the contributor.}\label{fig:atlas-interface}
\end{figure}
The studies suggest three distinct remedies. If a narrative omits a quality, a separate reporting channel may help. If a readout misses an available distinction, that operation requires evaluation. If a question turns an analogy into an object, a higher score can reinforce the wrong target. These are practical distinctions, not a universal theory of perception.

The atlas makes accounts, questionnaire responses, annotations, and images independently inspectable. Its 3,145 accounts support text search, vividness/duration filters, and comparison; the image collection links all 84 outputs to 24 accounts. Figure~\ref{fig:atlas-interface} shows the actual offline interface. A reader can search for an account, open its illustration record, and inspect the original wording, evidence status, and three prompt conditions together. In the displayed example, a recognizable beach does not by itself convey that the account describes a memory. Appendix Figure~\ref{fig:pair} separately compares two accounts with similar content but different vividness and duration. The record view is available at \texttt{atlas/browser/illustration.html} in the accompanying code. A separate blinded interface enables independent ratings; none have been collected. Appendix~\ref{app:atlas-views} details these views; their communication benefit remains unmeasured.

The public induced-experience sample is single-source and self-selected, with heterogeneous viewing conditions and unknown contributor overlap. It is not a clinical cohort and supplies neither neural measurements nor endorsed image targets. Synthetic diagnostics cover only seven template families; probes and combined-input models are exploratory. The image pilot uses selected short accounts, one backend, AI-authored evidence, and one judge. Hidden tool processing and possible backend drift across generation phases are uncontrolled. Its metric omits many aspects of experience and hallucination. None of these analyses establishes a simulation of a disorder, a diagnostic predictor, or improved empathy.

The communication goal should be tested with contributors and intended readers: compare account-only with account-plus-illustration, measuring what readers understand correctly and what they incorrectly infer. Attractive images and recognizable objects cannot substitute for that evidence.

\section{Conclusion}
PRT links reported accounts to inspectable representations and illustrations. The real-data study shows that content prediction is an incomplete proxy for reported experience; controlled contrasts distinguish similarity from recoverability; the image pilot exposes limits of both structured prompting and its evaluation. Keeping source meaning and interpretation visible makes these limits reviewable. Contributor endorsement and reader understanding remain the next substantive outcomes to test. 

A bridge between experiences begins with something another person can recognize, while leaving room for what only the speaker can clarify. Its value lies in making an unfamiliar experience discussable, with the person describing it still able to say what the image gets wrong.
\label{sec:main-end}
\clearpage
\section*{Reproducibility statement}
The accompanying code includes source checksums, selection and split rules, model revisions, exact generation prompts, all outputs, scoring questions and responses, analysis scripts, and figure generation. Protocol and request hashes were fixed before image generation. The source data are available at \url{https://osf.io/6dvh9/}. No new participant data were collected. Appendix~\ref{app:repro} gives execution details.
\bibliography{refs}
\bibliographystyle{iclr2027_conference}
\clearpage
\appendix

\section{Data provenance and label mappings}\label{app:data}
The input is \texttt{Ganzflicker\_Experience6k.csv}, OSF file \texttt{614a20b147d8290018671c0b}, version 1. Its SHA-256 is \texttt{0a716e983ea4\allowbreak{}b49d0df83af3\allowbreak{}7c6e33c229dd\allowbreak{}6b5bb9c4308b\allowbreak{}78bffc28322c\allowbreak{}eaca}. The original README is version 2 and also matches its OSF checksum. We retain original bytes and decode Windows-1252; UTF-8 decoding fails. All 6,665 logical records have 35 fields. Seven unnamed columns are empty. There are no entirely blank rows, repeated headers, or exact duplicate whole rows. These observations do not resolve the discrepancy with the documented count of 6,664.

The sequential eligibility exclusions are 90 records without explicit consent, 940 without an age in the specified range, and 2,490 outside the 5--400-word description range, leaving 3,145. These exclude neither all unusual answers nor all nonstandard viewing contexts. We do not imply reproduction of the original paper's 1,810-record final sample.

\begin{table}[h]
\centering\small
\begin{tabular}{lrrr}\toprule
Target & Train & Development & Test (positive)\\\midrule
Content & 1650 & 550 & 537 (144)\\
Vividness & 1479 & 483 & 474 (180)\\
Persistence & 1406 & 458 & 450 (281)\\
Affect & 1897 & 629 & 619 (234)\\\bottomrule
\end{tabular}
\caption{Valid task-specific record counts in the primary split. Missing targets are not imputed.}\label{tab:counts}
\end{table}

\paragraph{Content.} The allowed selections are distortions, simple shapes/patterns, colors other than red and black, complex objects, and complex environments, using the full exact strings in the source. Positive means one of the last two options is selected. Negative means a valid other selection or exactly ``No.'' Residual free text and combinations containing ``No'' plus other options are excluded rather than reinterpreted.
\paragraph{Vividness.} The ordered source options are weak/faint/insubstantial; clear but not vivid; clear and moderately vivid; clear/vivid/bright; and very vivid/almost real/popping out. Only an exact single option is valid. The last three are positive.
\paragraph{Persistence.} The ordered options are brief moment/flash; 1--2 seconds; persisting several seconds; and constant, morphing from one image to the next. Only an exact single option is valid. The last two are positive. This is a coarse threshold, not an assertion that constant morphing and several seconds are identical experiences.
\paragraph{Affect.} The two unpleasant options are positive. Neutral, somewhat enjoyable, and enjoyable are negative. Nonstandard strings are excluded. The target concerns the session's emotional rating, which may include aspects not described in the visual account.

Normalization for split grouping lowercases and collapses whitespace; it does not remove punctuation or identify paraphrases. The SHA-256 of \texttt{prt-study-v1:} concatenated with the normalized-text hash determines the primary split. Twelve inspected pilot groups are forced to training. For temporal sensitivity, groups are sorted by latest source-row occurrence; the first 60\% train, next 20\% develop, and last 20\% test. Pilot groups are excluded there. This avoids forcing late inspected examples into temporal training. Source timestamps are monotone but may reflect collection and questionnaire changes.

\section{Additional metrics and sensitivity}\label{app:metrics}
\begin{table}[h]
\centering\small\begin{tabular}{llrrr}\toprule
Task & Model & AUROC (95\% interval) & Balanced acc. & Brier\\\midrule
Content & TF--IDF & 0.877 [0.843, 0.910] & 0.590 & 0.142\\
Content & Imagery + context & 0.742 [0.699, 0.783] & 0.516 & 0.173\\
Content & MiniLM & 0.894 [0.861, 0.924] & 0.758 & 0.108\\
Content & MPNet & 0.891 [0.859, 0.922] & 0.769 & 0.112\\
Content & CLIP text & 0.887 [0.849, 0.921] & 0.771 & 0.107\\
Vividness & TF--IDF & 0.668 [0.617, 0.719] & 0.500 & 0.235\\
Vividness & Imagery + context & 0.740 [0.694, 0.787] & 0.645 & 0.197\\
Vividness & MiniLM & 0.645 [0.591, 0.696] & 0.573 & 0.229\\
Vividness & MPNet & 0.674 [0.624, 0.723] & 0.611 & 0.225\\
Vividness & CLIP text & 0.650 [0.595, 0.699] & 0.580 & 0.231\\
Persistence & TF--IDF & 0.652 [0.595, 0.705] & 0.500 & 0.231\\
Persistence & Imagery + context & 0.557 [0.502, 0.610] & 0.521 & 0.235\\
Persistence & MiniLM & 0.622 [0.569, 0.675] & 0.605 & 0.236\\
Persistence & MPNet & 0.640 [0.585, 0.689] & 0.601 & 0.233\\
Persistence & CLIP text & 0.605 [0.547, 0.661] & 0.602 & 0.240\\
Affect & TF--IDF & 0.543 [0.493, 0.586] & 0.500 & 0.235\\
Affect & Imagery + context & 0.579 [0.533, 0.626] & 0.522 & 0.232\\
Affect & MiniLM & 0.517 [0.472, 0.564] & 0.515 & 0.254\\
Affect & MPNet & 0.528 [0.482, 0.575] & 0.523 & 0.257\\
Affect & CLIP text & 0.519 [0.472, 0.567] & 0.515 & 0.256\\
\bottomrule\end{tabular}

\caption{Primary held-out metrics. Intervals resample known normalized-text groups and are conditional on fitted models. Balanced accuracy uses the unadjusted probability threshold 0.5.}
\label{tab:metrics}
\end{table}
\begin{table}[h]
\centering\small\begin{tabular}{llrrrr}\toprule
Split & Representation & Content & Vividness & Persistence & Affect\\\midrule
Primary & TF--IDF & 0.877 & 0.668 & 0.652 & 0.543\\
Primary & Imagery + context & 0.742 & 0.740 & 0.557 & 0.579\\
Primary & MiniLM & 0.894 & 0.645 & 0.622 & 0.517\\
Primary & MPNet & 0.891 & 0.674 & 0.640 & 0.528\\
Primary & CLIP text & 0.887 & 0.650 & 0.605 & 0.519\\
Temporal & TF--IDF & 0.878 & 0.639 & 0.616 & 0.536\\
Temporal & Imagery + context & 0.711 & 0.735 & 0.589 & 0.595\\
Temporal & MiniLM & 0.911 & 0.665 & 0.653 & 0.516\\
Temporal & MPNet & 0.889 & 0.682 & 0.623 & 0.508\\
Temporal & CLIP text & 0.903 & 0.663 & 0.621 & 0.539\\
\bottomrule\end{tabular}

\caption{AUROC under the primary and temporal group splits. These are two evaluations of the same source, not independent external replication.}
\label{tab:temporal}
\end{table}

The style baseline uses log word count, log character count, counts of commas, periods, question marks, and digits, and the type/token ratio. Context features contain imagery ratings and five recording-condition indicators; missing numeric values use the training median. Dense inputs are standardized from training data. TF--IDF uses unigrams/bigrams, minimum document frequency 3, maximum document frequency 0.98, sublinear term frequency, and at most 12,000 features. The masked variant removes a declared list of word stems associated with the target descriptions before fitting its own vocabulary. It does not remove all semantic synonyms.

The ten exploratory primary-split permutation runs independently shuffle training and development labels, select $C$ on shuffled development labels, and evaluate against unchanged test labels. Their AUROC ranges are 0.400--0.599 for content, 0.448--0.540 for vividness, 0.412--0.568 for persistence, and 0.469--0.533 for affect. These controls show finite-sample variation rather than certifying the absence of every form of leakage.

\section{Topic representation}\label{app:topics}
NMF uses 24 components, NNDSVDa initialization, random seed 20260912, and at most 400 iterations, fitted to training TF--IDF. The dimension is fixed rather than selected for an attractive visualization. Example high-loading terms include ``shapes, geometric shapes, geometric''; ``blue, yellow, green''; and ``light, balls, balls of light.'' Another component begins with ``but, were, they, there,'' demonstrating that a factor can capture reporting language rather than a perceptual category. Component labels are AI interpretations, not validated human coding. All 24 term lists are saved with the results.

\section{Synthetic contrasts and probe protocol}\label{app:contrasts}
Table~\ref{tab:templates} lists the anchor pairs. Each has an independently written faithful paraphrase; the full strings are included in the code. For example, certainty paraphrases are ``I am certain that $x$ appeared'' and ``I am uncertain whether $x$ appeared.'' Motion paraphrases are ``The $x$ kept changing position'' and ``The $x$ stayed completely still.'' The wrong option is either the other anchor or its paraphrase. The second form removes some asymmetric lexical-overlap advantage; it does not create a fully balanced natural-language benchmark.

\begin{table}[h]
\small\centering
\begin{tabular}{lp{.35\linewidth}p{.35\linewidth}}\toprule
Operator & Anchor 0 & Anchor 1\\\midrule
Certainty & I definitely saw $x$. & I might have seen $x$, but I am not sure.\\
Negation & I saw $x$. & I did not see $x$.\\
Duration & I saw $x$ continuously. & I saw $x$ for just an instant.\\
Motion & The $x$ were moving. & The $x$ were stationary.\\
Location & I saw $x$ in the center of my view. & I saw $x$ at the edges of my view.\\
Agency & I could make $x$ appear at will. & I could not make $x$ appear at will.\\
Seeing/imagining & I saw $x$ on the screen. & I imagined $x$ in my mind.\\\bottomrule
\end{tabular}
\caption{AI-authored template anchors. Contents are plural noun phrases. These statements are synthetic and not attributed to participants.}\label{tab:templates}
\end{table}

The 24 contents include colors and simple forms as well as faces, animals, buildings, trees, flowers, windows, ribbons, tunnels, clouds, rings, shadows, and human figures. We avoid inherently moving nouns in stationary templates. A content-control distractor preserves the query form but substitutes the content 12 positions away in the fixed list. Thus it changes both object identity and sometimes appearance; it is not matched to an operator change in edit distance or perceptual magnitude.

Faithful-choice intervals resample the 24 content families 2,000 times, keeping their polarity trials together. They quantify variation across the selected contents, not across new template families. We emphasize per-operator results rather than treating every template-generated trial as an independent observation.

For polarity probes, four consecutive six-content blocks rotate through test and development roles; the other two blocks train. Training uses anchors, whereas development and test use paraphrases. Each operator has its own standardized logistic probe with the same four $C$ values. This probes generalization across content and wording, but not to an unseen operator or human-authored paraphrase. Its accuracy is not directly comparable to unsupervised two-candidate retrieval, which uses different inputs and supervision.

\section{Development demonstrations}\label{app:development}
\begin{figure}[t]
\centering\includegraphics[width=\linewidth]{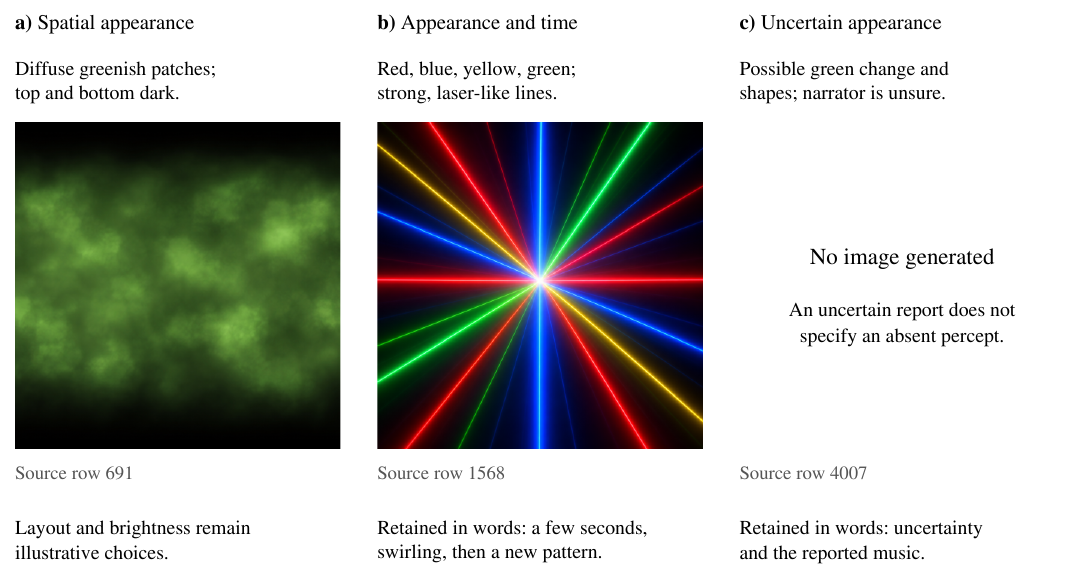}
\caption{\textbf{From reported attributes to a communication record.} (a) Source row 691 reports diffuse greenish patches and dark upper/lower boundaries. (b) Row 1568 describes a colored laser-like pattern, lasting seconds and changing; the still image illustrates appearance while the text retains time course. (c) Row 4007 expresses uncertainty about green coloration and shapes, so PRT retains the account without generating a definitive visual depiction. Descriptions above the outputs are condensed interpretations, not verbatim quotes. Images are generated development examples, not participant-endorsed targets.}
\label{fig:prt-demo}
\end{figure}

A preliminary compiler was run on three purposively selected development records from the existing pilot, covering spatial appearance, temporal change, and uncertainty (Figure~\ref{fig:prt-demo}). These induced visual experiences provide real accounts for demonstrating the workflow, without assigning a psychiatric or neurological diagnosis. Full source text, annotations, compiled prompts, and outputs are retained in the atlas demonstration. This development compiler used a stricter rendering policy, including abstention for an account without affirmative visual attributes. It is archived separately; the controlled comparison uses the policy in Section~\ref{sec:prt} and generates an image for every condition.

The process reveals an annotation failure as well as a possible use. For row 1568, the initial pilot record contained colors but omitted ``strong lines.'' The first image therefore showed colored shapes. Adding the missing lines still produced unsupported circles and triangles; preserving the source's ``laser lightshow'' description yielded the displayed illustration. All three attempts are archived. Row 691 used one generation call; row 4007 used none. This is an iterative demonstration, not a first-attempt success rate or a controlled comparison of generators.

The resulting record lets a reader inspect a visual interpretation alongside its source and the temporal or uncertain aspects it cannot depict. Exact layout and brightness remain rendering choices. We have not yet tested whether this record improves another person's understanding, or whether contributors prefer it to their account alone. That communication outcome is the relevant next test, rather than whether a model recognizes the intended objects.

\section{Atlas schema and evidence boundaries}\label{app:atlas}
\paragraph{Atlas implementation and clinical translation.}
An experience atlas should let a reader find accounts with a particular feature, inspect their source, and compare qualities that a single diagnosis or similarity score would hide. We implement a local descriptive prototype containing the 3,145 eligible accounts, with text search, vividness and duration filters, and side-by-side comparison of original content, vividness, duration, and feeling responses. It contains induced experiences, not a clinical cohort. We use \emph{atlas} for this organized collection; the current prototype establishes neither a taxonomy of disorders nor a validated map of subjective experience.

An experience atlas should be useful for communication, not merely recognizable to a model. The first design implication is to preserve separate fields for visual content, time course, uncertainty, context, and interpretation, with evidence spans and unknown values. A description of uncertain green coloration should not automatically become a confidently green scene. A narrative about visual static should not automatically acquire an auditory component because it uses a sound-related analogy.

The second implication is to evaluate the operation actually used. If an atlas retrieves neighboring reports by cosine similarity, a classifier's success is insufficient validation. If a tool generates illustrations, text-encoder accuracy is insufficient validation. A future reconstruction study should ask contributors what a depiction gets right, gets wrong, or cannot represent, then test whether it improves another person's understanding of the reported difficulties. The current experiments establish neither participant endorsement nor such communication benefit.

For clinical translation, the next decisive test is whether contributors endorse an illustration and whether another reader understands the reported difficulty more accurately from the account plus illustration than from the account alone. This would assess communication rather than presumed access to a private percept.

The third implication is to preserve the distinction between accounts, questionnaire responses, annotations, and generated media. PRT should support communication while keeping each interpretation traceable to its evidence.

\begin{figure}[h]
\centering\includegraphics[width=\linewidth]{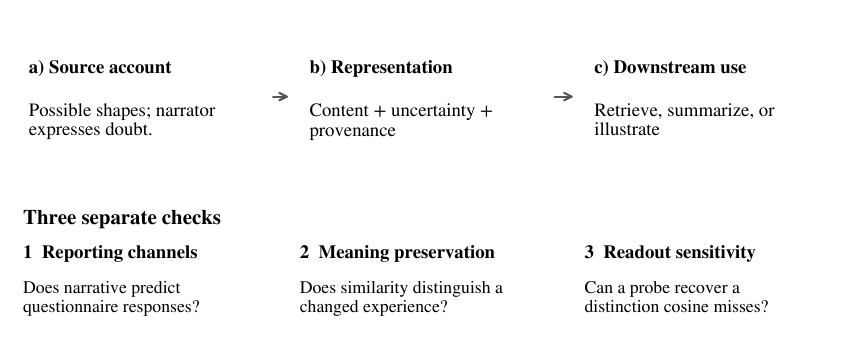}
\caption{\textbf{Three checks before an illustration is trusted.} (a) Source account, (b) representation, and (c) downstream use, with separate checks for reporting channels, similarity ordering, and readout sensitivity. It is a design aid, not a validated model of subjective experience.}\label{fig:framework}
\end{figure}
The development pilot contains 12 deterministically selected original accounts and 47 AI-authored annotations. Each annotation includes its interpretation, reported/uncertain/negated status, and an exact source-text span. Human annotations are a separate empty field. No participant has endorsed these annotations or the generated illustrations. The pilot is not a clinical corpus or a gold-standard test set.

Atlas records should distinguish a source account, optional diagnostic context explicitly supplied by a contributor, researcher annotations, model annotations, and generated media. They should retain reuse terms and the provenance of any pairing. Unknown attributes and aspects that cannot be depicted should remain visible. The browser reproduces source responses rather than replacing them with binary study labels or model annotations. It omits demographic fields and does not infer diagnoses. The present empirical analysis uses original questionnaire answers, not pilot annotations, and does not train a report-to-image generator. Figure~\ref{fig:visual} is an author-specified illustration of the communication goal. Figure~\ref{fig:prt-demo} uses source-linked pilot records; generation prompts, annotation edits, and all image attempts are archived.

The MUSE publication is methodological background only: its data-sharing statement says participants did not agree to dataset sharing \citep{melvin2021}. A public image-recreation deposit was identified but its dataset reuse terms and record pairing were not established in this audit. The current MOSAIC repository supplies analysis code but no raw accounts in its checked data folders. The source audit records these distinctions without presenting all readable publications as available datasets.

\section{Execution and reproducibility}\label{app:repro}
Experiments run locally on Apple silicon with 64~GB memory, Python 3.11.13, PyTorch 2.8.0, Transformers 4.56.1, Sentence Transformers 5.1.0, and scikit-learn 1.7.1. Neural inference uses the MPS backend. Models remain frozen; logistic probes are fitted locally. The numerical encoder experiments use local inference. The development demonstration sends selected source-linked visual excerpts to the image-generation backend; the controlled comparison sends the full selected accounts with condition-specific instructions.

Model identifiers are \path{sentence-transformers/all-MiniLM-L6-v2}, \path{sentence-transformers/all-mpnet-base-v2}, and \path{openai/clip-vit-base-patch32}. 
Output dimensions are 384, 768, and 512. Checkpoint cards, source metadata, derived-file checksums, and execution logs accompany the local research package.

The execution order is source retrieval, eligibility/target preparation, frozen encoding, primary and temporal prediction, synthetic contrasts, exploratory follow-ups, and figure/table generation. Source files remain unchanged. A separate protocol distinguishes the initial plan from exploratory additions. Tests check target mappings, split separation, embedding alignment, stored predictions, source checksums, and evidence provenance. They establish computational consistency, not human validation of the conceptual framework. Reproduction instructions and limitations accompany the code.

\section{Controlled generation and scoring details}\label{app:generation}
The fixed protocol, 24 source records, 74 annotations, 69 questions, and 144 shuffled requests are stored separately. The original 144-call plan was amended for budget reasons after 62 outputs. Twenty-two additional calls completed the 24 three-condition groups; the earliest output per condition supplies the 72-image primary analysis. Both evidence conditions include each annotation's status, exact source span, and interpretation. Prose and structured prompts reproduce the same evidence; tests verify byte-for-byte agreement between the active compiler and the frozen requests. The original account remains verbatim in all conditions. Source-span checks reject unmatched evidence, but cannot validate an interpretation.

For illustration scoring, Qwen2.5-VL-3B-Instruct uses revision \texttt{66285546d2b8\allowbreak{}21cf421d4f5e\allowbreak{}b2576359d377\allowbreak{}0cd3}, half precision on MPS, eager attention, deterministic decoding, and a maximum of 12 new tokens. Image processing uses the installed fast processor, a minimum of 256 and maximum of 512 visual tokens' worth of pixels ($28^2$ pixels per token). The exact prompt asks for YES, NO, or UNCERTAIN based only on visible content; neither condition labels nor expected answers enter the judge. Questions are shuffled with a fixed seed. Full raw responses are retained, including parsing failures.

Confirmed coverage weights reports equally after averaging their supported-attribute questions in the earliest output per condition. An upper bound permits uncertain or unparseable responses to count as present; this is retained with the results rather than silently recoding them as absence. The three unsupported-literalization questions concern an actual coffee vessel, train hardware, and anatomically depicted blood vessels. They are not representative of every unsupported object a generator could add.

\section{Prediction protocol and supporting analyses}\label{app:prediction-protocol}
\subsection{Source, selection, and targets}
We use the public Ganzflicker deposit associated with \citet{reeder2022}. Appendix~\ref{app:data} documents source checksums, eligibility rules, and the unresolved difference between 6,665 parsed records and the source's documented 6,664.

We retain records with explicit consent, a numeric reported age from 18 to 100, and a visual description containing 5--400 whitespace-delimited words. This leaves 3,145 records and 3,134 distinct normalized texts. These rules define our analysis sample; they do not reproduce the original study's observation-condition exclusions. The data include heterogeneous viewing conditions, self-reported attributes, and no validated stable person identifier. We therefore use ``records,'' not a claim that all retained rows identify distinct people.

The questionnaire provides four targets. \emph{Content} indicates selection of complex objects or environments versus valid selections without these options. \emph{Vividness} indicates an exact single answer at least ``clear and moderately vivid.'' \emph{Persistence} indicates an exact single answer of several seconds or constant/morphing, versus a flash or 1--2 seconds. \emph{Affect} indicates an unpleasant rating versus neutral or enjoyable. Missing, mixed, contradictory, or nonstandard answers are excluded from the affected task only. This deliberately avoids converting ambiguous multi-selections into a single severity score. Appendix~\ref{app:data} specifies the mappings and counts.

\subsection{Splits, models, and metrics}
Normalized-text hashes assign groups to training, development, and test in approximately 60/20/20 proportions: 1,897, 629, and 619 records. Exact text duplicates cannot cross splits. The previously inspected pilot groups are forced to training. Different surface forms from the same unidentifiable contributor could still cross splits; this is a limitation of the source. The primary test sets contain 537, 474, 450, and 619 valid records for the four targets respectively.

We compare a prevalence baseline, seven length/style features, seven imagery/context covariates, word TF--IDF, a 24-component nonnegative factorization of training TF--IDF \citep{lee1999}, and frozen MiniLM, MPNet, and CLIP text encoders. The context covariates contain visual and auditory imagery ratings and indicators for darkness, computer use, white noise, and viewing duration. Neither the target questionnaire answer nor diagnosis enters a narrative representation. The context model addresses a different input channel and is not advertised as a text-only competitor.

For comparable narrative coverage, all neural models encode the same non-overlapping 32-word chunks. We normalize chunk embeddings, average them equally, and normalize the result. Two of 5,611 chunks exceed CLIP's 77-token limit and are truncated; none exceed the sentence encoders' limits. The operator diagnostic uses short sentences and does not involve multi-chunk pooling. Exact model revisions and software are recorded in Appendix~\ref{app:repro}.

Each feature representation uses a logistic head with $C\in\{0.01,0.1,1,10\}$ selected by development AUROC. Transformations are fitted on training data only; the chosen model is not refitted on the test set. We report AUROC (how well scores rank positive above negative responses; chance is 0.5), balanced accuracy at threshold 0.5, and Brier score (probability error; lower is better). Confidence intervals resample normalized-text groups 500 times, conditional on the fitted model. Paired differences use 1,000 resamples. These intervals do not account for unknown contributor dependence or every source of model-selection uncertainty.

The analysis plan was written before fitting, after the schema pilot. It is an internal prospective plan, not a public preregistration. Chronological sensitivity, lexical masking, and label permutation check specific alternative explanations. Combined-input models and polarity probes were added after viewing initial results and are explicitly exploratory.

\subsection{Full primary comparison and context sensitivity}
\begin{figure}[t]
\centering\includegraphics[width=\linewidth]{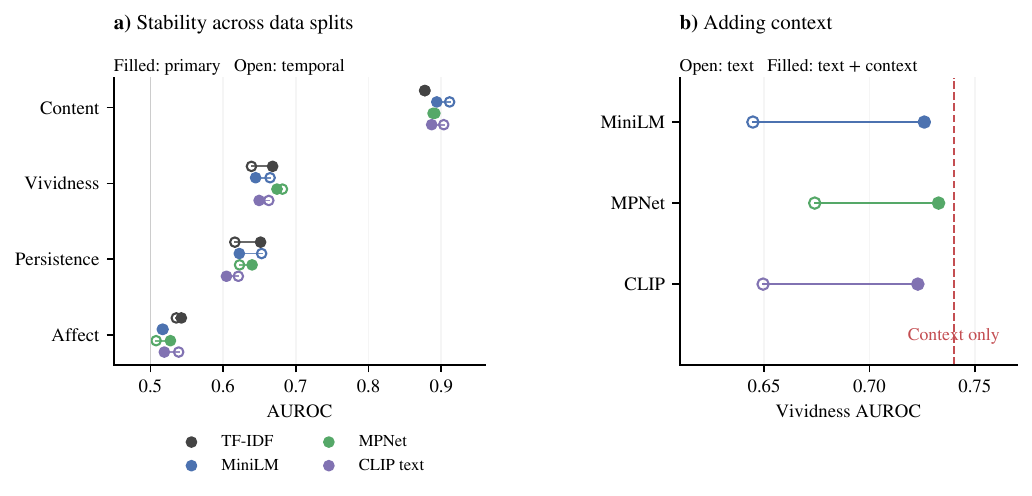}
\caption{\textbf{The result depends on the question and available inputs.} (a) The broad content-versus-other-dimensions pattern persists under a time-ordered group split, although rankings change. (b) Exploratory addition of imagery/context inputs improves vividness prediction relative to narrative embeddings alone; it does not outperform the context-only point estimate. Intervals and paired comparisons are supplied with the results.}
\label{fig:sensitivity}
\end{figure}
\begin{table}[t]
\caption{\textbf{Primary held-out AUROC.} All models use the same task-specific records. The context row has additional inputs. Masked TF--IDF removes a prespecified set of explicit content, vividness, duration, and affect words. Full intervals and additional metrics appear in Appendix~\ref{app:metrics}.}
\label{tab:primary}
\centering\small\begin{tabular}{lrrrr}\toprule
Representation & Content & Vividness & Persistence & Affect\\\midrule
Prevalence & 0.500 & 0.500 & 0.500 & 0.500\\
Length/style & 0.651 & 0.591 & 0.616 & 0.524\\
Imagery + context & 0.742 & 0.740 & 0.557 & 0.579\\
TF--IDF & 0.877 & 0.668 & 0.652 & 0.543\\
NMF (24 topics) & 0.804 & 0.599 & 0.621 & 0.487\\
MiniLM & 0.894 & 0.645 & 0.622 & 0.517\\
MPNet & 0.891 & 0.674 & 0.640 & 0.528\\
CLIP text & 0.887 & 0.650 & 0.605 & 0.519\\
Masked TF--IDF & 0.859 & 0.651 & 0.636 & 0.545\\
\bottomrule\end{tabular}

\end{table}\paragraph{Additional input matters more than a blanket encoder upgrade.}
The imagery/context model predicts vividness at 0.740 AUROC. In the exploratory concatenation experiment, narrative-plus-context AUROC is 0.733 for MPNet and 0.723 for CLIP, while MiniLM also improves over its narrative-only version. MPNet-plus-context differs from context alone by $-0.007$ $[-0.050,0.039]$. Thus adding context helps relative to the corresponding narrative model, but this experiment does not establish an improvement over context alone. Figure~\ref{fig:sensitivity} makes both comparisons explicit. Known associations between imagery and induced perceptual vividness \citep{reeder2022} provide a plausible explanation; this is not a novel psychiatric predictor.

\paragraph{Sensitivity checks limit stronger interpretations.}
A temporal group split retains high content prediction (neural AUROC 0.889--0.911) and weak affect prediction (0.508--0.539). Removing explicit task-related words reduces lexical content AUROC from 0.877 to 0.859. Ten label-permutation runs show substantial finite-sample variation around chance. These within-source checks address specific alternatives, not external replication or causal identification; Appendix~\ref{app:metrics} gives details.

\begin{figure}[h]
\centering\includegraphics[width=.85\linewidth]{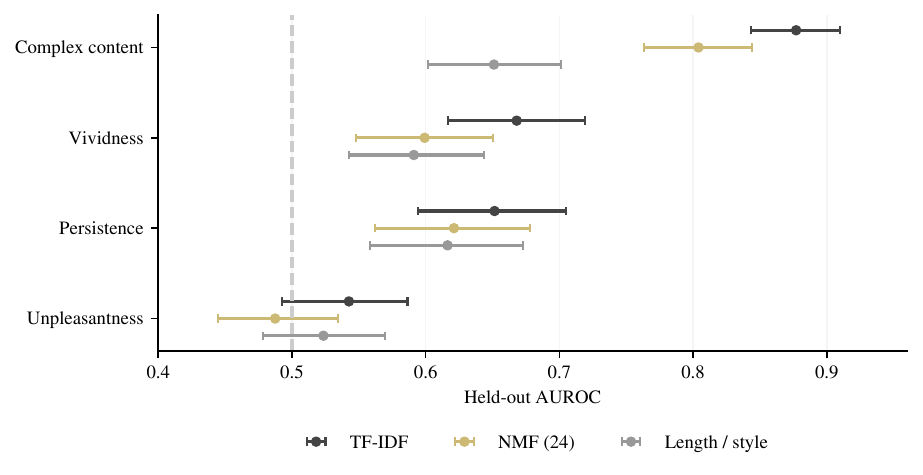}
\caption{\textbf{A compact lexical decomposition.} Predictive performance of 24 NMF components versus full TF--IDF and style features. The representation retains some content information but does not establish a universal atlas of experiences.}\label{fig:topics}
\end{figure}

\paragraph{Topic compression retains content better than affect.}
A 24-component NMF representation reaches 0.804 content AUROC but 0.487 for affect. Components include lexical groups around geometric shapes, color, light, and the screen, as well as a component dominated by common reporting words. Figure~\ref{fig:topics} compares the topic representation with lexical and style features; Appendix~\ref{app:topics} gives representative loadings. These components are interpretable descriptions of word co-occurrence, not established perceptual primitives or biological subtypes. A visually attractive map of them would not establish those stronger claims.

\section{Readout probes and their interpretation}\label{app:readout}
\subsection{Exploratory probes recover some distinctions}
We next fit a separate binary polarity probe for each operator using frozen representations. A probe is a simple trained classifier that reads a distinction from otherwise unchanged encoder outputs. In four folds, 12 contents provide anchor-wording training examples, six provide paraphrase development examples, and six provide paraphrase test examples. Each content appears in the test set once, producing 48 held-out decisions per operator. The operator family is known; this is not transfer to a new semantic operation.

Figure~\ref{fig:readout} shows recoverability that varies by operator and model. Averaged over the seven operators, polarity accuracy is 83.9\% for MiniLM, 93.5\% for MPNet, and 64.0\% for CLIP. Recovery is not uniform: CLIP's motion probe remains below chance at 37.5\%, while its agency probe reaches 81.3\%. Thus a linear probe is a diagnostic tool, not a universal correction. Wording generalization, small content families, and operator-specific supervision constrain the result.

These observations support a careful conclusion: some similarities are poor proxies for distinctions that another readout can access. They do not show that a generator will preserve those distinctions, that a probe captures a neural mechanism, or that all failures can be fixed with a linear layer. Existing work already distinguishes encoded information from alignment and readout behavior \citep{koishigarina2026,lu2026}; our analysis brings that distinction into an experience-report audit.

\paragraph{Why the distinction matters.}
Similarity and recoverability answer different practical questions. A nearest-neighbor search can return an account with the wrong qualification even when a trained classifier can recover that qualification from the same representation. For example, $z=(u,\epsilon a)$ can retain an attribute $a$ in a small coordinate while cosine is dominated by content $u$. This explains a possible discrepancy, not a proven mechanism of our encoders. A communication atlas should evaluate its actual retrieval or generation operation rather than borrow reassurance from another task's score.

\section{Illustration protocol and complete numerical results}\label{app:image-protocol}
\subsection{Reports, annotations, and generation}
The underlying source contains visual descriptions and separate questionnaire responses. We retain adult records with explicit consent and descriptions of 5--400 words, giving 3,145 records and 3,134 unique normalized texts. Normalized-text groups are assigned to training, development, and test; the earlier 12-record development pilot belongs to training. Appendix~\ref{app:data} gives eligibility, target mappings, and source provenance.

For the illustration study, we select 24 unique test groups with descriptions of 12--70 words, ordered by a fixed salted text hash. Selection precedes annotation and image generation. There are 74 source-linked annotations and 69 scoring questions. Of these, 66 concern supported visible details; three separately check specific unsupported literalizations. The latter are deliberately narrow and do not constitute a comprehensive hallucination measure.

The initial frozen plan called for two images in each condition (144 calls). After 62 outputs, generation was paused for budget reasons. A documented amendment authorized exactly 22 missing-condition calls, completing at least one image for every report and condition. The primary analysis uses the earliest recorded output for each report--condition pair: 72 images from all 24 original reports. Selection never uses output quality. The 12 additional outputs are retained for a sensitivity analysis. Prompts, evidence, and scoring questions remain unchanged; this is a budget-amended pilot, not the original repeated-output design. Earlier iterative development cases are excluded (Appendix~\ref{app:development}).

The backend is the built-in image-generation tool; embedded output metadata identifies \texttt{gpt-image}, version 2.0. We save every returned image, prompt hash, source path, and available model metadata. The interface does not expose an exact weights revision, random seed, or internal prompt processing. Our design therefore matches the number of generation calls, not identical latent draws or compute. The source prompts request one square image and permit reported symbols while excluding added explanatory captions.

\subsection{Measurements and uncertainty}
A separate local Qwen2.5-VL-3B-Instruct model \citep{qwen2025vl} answers fixed questions about each image, without condition labels, original generation prompts, or expected answers. Questions concern visible properties such as a red ball, a grid, or an upper-left position. The judge returns yes, no, or uncertain. It answers all 32 deterministic color, shape, object-absence, and position fixture questions correctly. These elementary checks establish basic pipeline behavior, not accuracy on complex generated scenes.

The primary metric is \emph{confirmed attribute coverage}. For each report and condition, we average the fraction of supported attributes answered yes for the earliest output in that condition; we then average reports equally. Uncertain and unparseable answers are not counted as confirmed, and their counts are retained. Unsupported-literalization probes are reported separately. The questions do not score motion, duration, emotional experience, or whether the original contributor would endorse an image.

For report $i$, let $c_{i,S}$ and $c_{i,E}$ denote those coverage scores. The primary effect is
\begin{equation}
\widehat\Delta_{S-E}=\frac{1}{24}\sum_{i=1}^{24}(c_{i,S}-c_{i,E}).
\end{equation}
Confidence intervals use 10,000 paired bootstrap resamples of reports; two-sided paired randomization tests use 100,000 sign flips. The unit of comparison is the report: 24 accounts, not 72 independent people. Intervals remain conditional on this source sample, annotation procedure, image backend, and automated judge.

\begin{table}[t]
\centering\small
\caption{\textbf{Controlled illustration results.} Coverage averages supported-attribute judgments within report before averaging 24 reports. The last column counts confirmed unsupported literalizations over three images per condition, from only three specific source accounts. Lower is better for that column; it is not a general hallucination rate.}
\label{tab:controlled}
\begin{tabular}{lcc}
\toprule
Prompt & Coverage [95\% interval] & Literalizations \\
\midrule
Original report & 92.4\% [81.9, 100.0] & 0/3 \\
Evidence in prose & 100.0\% [100.0, 100.0] & 0/3 \\
PRT: structured evidence & 97.9\% [93.8, 100.0] & 0/3 \\
\bottomrule
\end{tabular}

\end{table}
\section{Atlas views and comparison examples}\label{app:atlas-views}
\begin{figure}[t]
\centering\includegraphics[width=\linewidth]{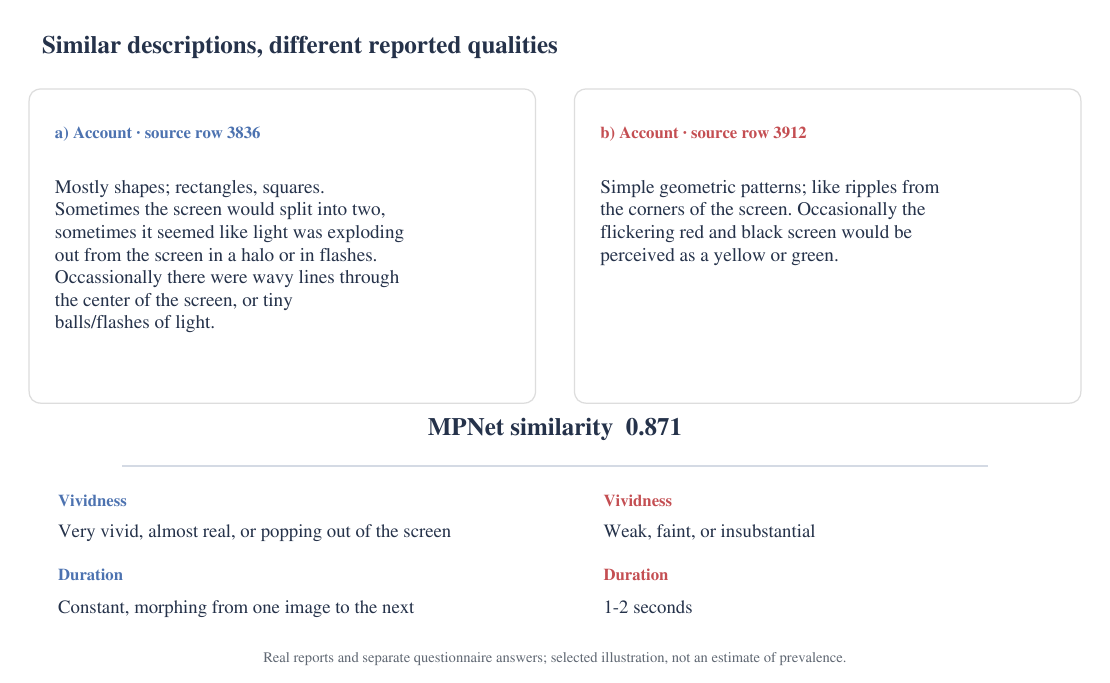}
\caption{\textbf{Similar content, different reported qualities.} Designed example, not an interface screenshot: (a, b) source accounts with their separate questionnaire responses. The pair was selected post hoc for high MPNet similarity among eligible training accounts with matching content category but different vividness and persistence labels. It illustrates a distinction an atlas should expose; it is not an estimate of how often retrieval makes this error.}\label{fig:pair}
\end{figure}

An atlas should help someone find a relevant account, understand what is similar and different, and inspect how an illustration was derived. Our offline prototype contains 3,145 source accounts, text search, vividness and duration filters, and side-by-side comparison of original questionnaire responses. The controlled image collection links 84 retained outputs to 24 accounts, with 72 images in the primary matched analysis. A separate blinded interface presents images and source questions for future independent review; no human ratings have been collected.

The account remains the common reference across these views. Study~1 motivates retaining separate reporting channels. Study~2 motivates checking the similarity operation rather than assuming an embedding's probe accuracy guarantees useful retrieval. Study~3 motivates preserving the source span and qualification behind each visual instruction and scoring question. These are design implications of the audits, not measured improvements in user understanding.

Figure~\ref{fig:pair} illustrates the first two concerns using two source accounts. Their content similarity does not make their separately reported vividness and duration interchangeable. Appendix~\ref{app:atlas} documents the schema, and Appendix~\ref{app:development} retains the earlier source-linked development illustrations and the uncertainty-only record. The development compiler's abstention policy was not used in the matched image study.

\paragraph{Study overview.}
Table~\ref{tab:roadmap} summarizes the different roles of the three studies.
\begin{table}[t]
\centering\small
\caption{\textbf{Three connected questions, three kinds of evidence.} The studies evaluate different parts of the same communication workflow. No single result validates the full path from an experience to another person's understanding.}\label{tab:roadmap}
\begin{tabular}{p{.22\linewidth}p{.33\linewidth}p{.36\linewidth}}\toprule
Question & Evidence & What the result informs\\\midrule
What is reported? & 3,145 public accounts; separate questionnaire responses & Which qualities a visual narrative makes predictable.\\
What meaning is preserved? & 672 controlled synthetic contrasts; exploratory probes & Whether wording and readout change access to a distinction.\\
What does an illustration convey? & 24 matched three-condition groups; automated image questions & Visible-detail coverage, prompt-baseline comparison, and limits of the criterion.\\\bottomrule
\end{tabular}
\end{table}
\end{document}